\documentclass[aps,prd,
preprintnumbers,groupedaddress]{revtex4}
\usepackage{amsmath}
\usepackage{bm}

\begin{document}
\preprint{\vtop{
{\hbox{YITP-13-130}\vskip-0pt
}
}
}


\title{Hidden-Strangeness Partners of $\bm{X(3872)}$
}
\author{
Kunihiko Terasaki   
}
\affiliation{
Yukawa Institute for Theoretical Physics, Kyoto University,
Kyoto 606-8502, Japan 
}

\begin{abstract}{
Decay properties of hidden-strangeness partners of ${X(3872)}$ are studied in tetra-quark pictures.  
As the result, it is seen that their decay properties are strongly model dependent, and therefore,  
experimental studies of them will select a realistic model. 
} 
\end{abstract}

\maketitle

Recently hidden-charm axial-vector mesons $Z_c^{\pm,0}(3900)$ have been discovered in $\pi J/\psi$ 
channel from $e^+e^-\rightarrow \pi\pi J/\psi$~\cite{BESIII,Belle-Z(3895),Xiao}. 
($J/\psi$ is written as $\psi$ hereafter.) 
Existence of such states was predicted from tetra-quark pictures~\cite{Maiani-X,X-3872-KT} 
much earlier than the discovery, 
and after the observation, they have been interpreted as partners of  $X(3872)$ from various 
pictures~\cite{Z_c-models,Guo,Voloshin}, and the above predictions have been 
updated~\cite{Maiani-odd-C,Z_c-3900-KT}. 
If they are $D\bar{D}^*$ molecules~\cite{Guo}, experiments would have observed two peaks arising 
from $X(3872)$ and $Z_c^0(3900)$ in $D^0\bar{D}^{*0}$ and its charge-conjugate ($c.c.$) channels. 
However, no signal of $Z_c^0(3900)$ was observed in $D\bar{D}^{*} \oplus c.c.$ channels in  
Belle~\cite{Belle-X-3875} and Babar~\cite{Babar-X-3875} experiments. 
In addition, there are some arguments that $X(3872)$ as a compact object is 
favored~\cite{prompt-X-theor,TWQCD}. 
If the above arguments are the case, $X(3872)$ and $Z_c(3900)^{\pm,0}$ would be tetra-quark mesons. 
However, these two tetra-quark models in \cite{Maiani-X} and \cite{X-3872-KT} are very much 
different from each other. 
To see more explicitly such differences, we study hidden-charm and -strangeness axial-vector 
tetra-quark mesons. 
Such mesons would be observed in $\phi\psi$ channel if kinematically allowed (if not, in $\gamma\psi$ 
and in $\gamma\phi$) when their charge conjugation ($\mathcal{C}$) parity is even ($\mathcal{C}=+$), 
and in $\eta\psi$ channel when $\mathcal{C}=-$. 
However, experimental results on $\phi\psi$ resonances are still controversial~\cite{phi-psi-exp}, and 
no narrow resonance signal has been observed~\cite{Miyabayashi} in $\eta\psi$ channel. 

Tetra-quark states can be classified into four groups as  
\begin{eqnarray} 
&&\hspace{-8mm} \{qq\bar q\bar q\} =  
[qq][\bar q\bar q] \oplus (qq)(\bar q\bar q)  \oplus \{[qq](\bar q\bar q)\oplus (qq)[\bar q\bar q]\}, 
                                                                                                                   \label{eq:4-quark} 
\end{eqnarray} 
where parentheses and square brackets denote symmetry and anti-symmetry, respectively, of flavor 
wave functions (wfs.)~\cite{Jaffe}.   
($q = u,\,d,\,s$ and $c$ in this note.)
Each term on the right-hand-side (r.h.s.) of Eq.~(\ref{eq:4-quark}) is again classified into two  
groups~\cite{Jaffe} with 
${{\bar{{3}}}_c}\times{{3}_c}$ and ${{6}_c}\times {{\bar{{6}}_c}}$  
of the color $SU_c(3)$. 
Here, the former is considered as the lower lying state in heavy mesons~\cite{D_{s0}-KT,HT-isospin}.     
(The flavor $(qq)(\bar q\bar q)$ multiplet and the color ${{6}_c}\times {{\bar{{6}}_c}}$ states 
are not considered in this note.) 
Regarding their spin ($J$) and parity ($P$), the first term and the last two on the r.h.s. of 
Eq.~(\ref{eq:4-quark}) have $J^P=0^+$ and $1^+$, respectively, in the flavor symmetry  
limit~\cite{Z_c-3900-KT}. 
Because the flavor symmetry is broken in the real world, however, all the states on the r.h.s. of  
Eq.~(\ref{eq:4-quark}), in particular, even $[qq][\bar q\bar q]$ can have all of $J^P=0^+,\,1^+$ and 
$2^+$. 
Along with this line, the diquarkonium model~\cite{Maiani-X} in which a broken flavor symmetry 
(and even a broken isospin symmetry) were assumed was proposed. 
In this model, tetra-quark mesons are given by products of a diquark $[qq]_{S}$ and an antidiquark  
$[\bar{q}\bar{q}]_{S'}$, where $S$ and $S'$ are spins of $[qq]$ and $[\bar{q}\bar{q}]$, respectively, and  
$S,\,S'=0,\,1$.  
Thus, this model predicts existence of (i) two different types of scalar mesons,  
$[qq]_{0}[\bar{q}\bar{q}]_{0}$ and $[qq]_{1}[\bar{q}\bar{q}]_{1}$, 
(ii) three different types of axial-vector mesons, 
$[qq]_{0}[\bar{q}\bar{q}]_{1} \pm [qq]_{1}[\bar{q}\bar{q}]_{0}$ and $[qq]_{1}[\bar{q}\bar{q}]_{1}$,   
(iii) tensor mesons, $[qq]_{1}[\bar{q}\bar{q}]_{1}$, 
while (iv) no double charm mesons. 
Although, in this model, it is assumed that the isospin symmetry is broken, 
experiments~\cite{Belle-isospin} suggest that isospin symmetry works well in $X(3872)$ physics. 
In addition, no charged partner of $X(3872)$ has been observed~\cite{Babar-X-charged-partner}.  
Therefore, $X(3872)$ is considered as an iso-singlet state. 
Furthermore, this model predicts existence of two different types of hidden-charm scalar mesons 
as discussed above, and that the mass of the lower $[cn]_0[\bar{c}\bar{n}]_0$ state (called as $X_0$) is 
around 3.7 GeV, because $X_0$ and $X(3872)$ have been assigned to elements of multiplets with the  
same flavor structure. 
In contrast, an indication of $\eta\pi^0$ peak (called as $\hat{\delta}^{c0}(3200)$ in this note)  
has been observed around 3.2 GeV in $\gamma\gamma$ collision~\cite{Uehara}. 
This can be interpreted as $\{[cn]_0[\bar{c}\bar{n}]_0\}_{I=1}^0$ whose mass is expected to be 
$\simeq 3.3$ GeV (close to $m_{\hat{\delta}^{c0}(3200)}$), by using a quark 
counting~\cite{D_{s0}-KT,hidden-charm-scalar-KT} with the mass differences 
$\Delta_{sn}= m_s - m_n\simeq 0.1$ GeV and $\Delta_{cs}=m_c - m_s\simeq 1$ GeV, 
where the mass of $D_{s0}(2317)$ has been taken as the input data.  

The above interpretation is based on the alternative scheme that tetra-quark mesons with $J^P=0^+$  
and $1^+$ are assigned to different flavor multiplets, 
$[qq]_0[\bar{q}\bar{q}]_0$ and $\{[qq]_0(\bar{q}\bar{q})_1\oplus (qq)_1[\bar{q}\bar{q}]_0\}$,  
respectively~\cite{X-3872-KT,D_{s0}-KT,hidden-charm-scalar-KT}. 
In this scheme, double-charm axial-vector $(cc)_1[\bar{q}\bar{q}]_0$ mesons can 
exist~\cite{X-3872-KT,Exotic-QN}, in contrast with the diquarkonium model. 
It is true that the flavor symmetry is broken in the real world,  but the symmetry breaking in spatial 
wfs. seems to be not so bad even in $SU_f(4)$, as seen below. 
A matrix element of flavor charge taken between $\langle{\alpha}|$ and $|{\beta}\rangle$ are given by  
a corresponding form factor $f_+^{(\alpha\beta)}(0)$ of matrix element of vector current at zero  
momentum transfer squared. 
Thus, a deviation of spatial wave function overlap from the flavor symmetry limit will be given by 
a deviation of the corresponding $f_+^{(\alpha\beta)}(0)$ from unity. 
Therefore, it is significant to remember the values of form factors 
$f_+^{(\pi K)}(0) = 0.961 \pm 0.008$~\cite{Leutwyler},    
$f_+^{(\bar{K}D)}(0) = 0.74\pm 0.03$~\cite{PDG96},  
$f_+^{(\pi D)}(0)/{f_+^{(\bar{K}D)}(0)} = 
1.00 \pm 0.11 \pm 0.02$~\cite{FNAL-E687} and
$0.99 \pm 0.08$~\cite{CLEO-FF}. 
From these results, it is seen that spatial wfs. satisfy well the $SU_f(3)$ symmetry as has recently  
been confirmed~\cite{FF-SU-3-symm}, and that the $SU_f(4)$ symmetry breaking is not necessarily 
so bad in contrast with masses. 
Thus, we have assigned~\cite{D_{s0}-KT,hidden-charm-scalar-KT} scalar $D_{s0}(2317)$ and  
$\hat{\delta}^c(3200)$ to $\{[cn]_0[\bar{s}\bar{n}]_0\}_{I=1}$ and $\{[cn]_0[\bar{c}\bar{n}]_0\}_{I=1}$, 
respectively, as the $0$-th order approximation of the flavor symmetry breaking.   
In the case of the $J^P=1^+$ hidden-charm mesons, however, $[cq]_0(\bar{c}\bar{q})_1$ and  
$(cq)_1[\bar{c}\bar{q}]_0$ have no definite $\mathcal{C}$-property, so that they mix with each other in 
a way that their flavorless neutral components form eigenstates of $\mathcal{C}$-parity. 
Thus, axial-vector $X(3872)$ and $Z_c(3900)$ are assigned to 
$X(+)\sim\{[cn]_0(\bar{c}\bar{n})_1 + (cn)_1[\bar{c}\bar{n}]_0\}_{I=0}$ and 
$X_I(-)\sim\{[cn]_0(\bar{c}\bar{n})_1 - (cn)_1[\bar{c}\bar{n}]_0\}_{I=1}$, 
respectively~\cite{X-3872-KT,Z_c-3900-KT}. 
(In this short note, we always assume that mixings of flavorless states except for the $\eta\eta'$ 
mixing are ideal. This assumption is compatible with the OZI-rule~\cite{OZI}.) 
It should be noted that the mass difference between $X(+)=X(3872)$ and $X_I(-)=Z_c(3900)$ is small 
($\lesssim 1\,\,\%$)~\cite{Z_c-3900-KT}. 
This seems to suggest that spatial wfs. of $[cn](\bar{c}\bar{n}) \pm (cn)[\bar{c}\bar{n}]$ states are not  
so different from each other. 

As seen above, flavor structures of axial-vector mesons in these two models are drastically different 
from each other, and therefore, it is supposed that their decay properties also are very much different  
from each other. 

In accordance with the prescription in \cite{Jaffe}, we decompose each of 
$X^s(\pm)\sim 
\{[cs]_{\bar{3}_c}^{1_s}(\bar{c}\bar{s})_{3_c}^{3_s} 
\pm (cs)_{\bar{3}_c}^{3_s}[\bar{c}\bar{s}]_{3_c}^{1_s}\}_{1_c}^{3_s}$ 
in our tetra-quark model into a sum of products of $\{q\bar{q}\}$ pairs, and then replace color 
singlet pairs by ordinary mesons, for example, $\{c\bar{c}\}_{1_c}^{1_s}$ by $\eta_c$, 
$\{c\bar{c}\}_{1_c}^{3_s}$ by $\psi$, $\cdots$ (in the case of the low lying mesons), 
where the notation in \cite{KT-partners-of-X} has been taken to see explicitly color and spin 
degree of freedom, i.e., the subscript and superscript of $q$, $\bar{q}$ and their products denote their  
color and spin multiplets; the color ${3}_c$, $\bar{3}_c$, and $1_c$ of $SU_c(3)$, and the spin singlet  
$1_s$ and triplet $3_s$, respectively. 
The results are 
\begin{eqnarray}                    
&&\hspace{-5mm}
X^s(+)
= \frac{1}{2}\sqrt{\frac{1}{6}}\Bigl\{
\sqrt{2}[\psi\phi - \phi\psi] 
+ [D_s^+D_s^{*-} - D_s^-D_s^{*+} + D_s^{*+}D_s^- - D_s^{*-}D_s^{+}] 
\Bigr\} + \cdots,                               \label{eq:decomp-X^s_+}
\\
&&\hspace{-5mm}
X^s(-) 
= \frac{1}{2}\sqrt{\frac{1}{6}}\Bigl\{
[\eta_c\phi - \phi\eta_c] + [\psi\eta_s  -  \eta_s\psi ] 
+ \sqrt{2}[D_s^{*+}D_s^{*-} - D_s^{*-}D_s^{*+}]
\Bigr\} + \cdots,                                 \label{eq:decomp-X^s_-}
\end{eqnarray}
where $\eta_s= \{s\bar{s}\}_{1_c}^{1_s}$, and $\cdots$ denotes a sum of  
$\{q\bar{q}\}_{8_c}\{q\bar{q}\}_{8_c}$ products (and possible contributions of excited ordinary mesons). 
The above decompositions could be reshuffled by exchanging a gluon between intermediate quarks. 
However, we here neglect such a reshuffling, because a quark-gluon coupling is not very strong 
around the energy scale under consideration~\cite{PDG10}. 
Under this approximation, $X^s(+)$ with $\mathcal{C}=+$ couples to $D_s^+D_s^{*-} + c.c.$, as 
$X(3872) = X(+)$ couples to $D^0\bar{D}^{*0} + c.c.$~\cite{KT-partners-of-X}. 
(The decay $X(3872)\rightarrow D^0\bar{D}^{*0} + c.c.$ has actually been observed as a dominant 
mode.) 
However, the threshold of $D_s^+D_s^{*-}$ decay is 
a little bit higer than $m_{X^s(+)}\simeq 4.07$ GeV which is estimated by using the same quark 
counting as the above (but now with $m_{X(+)} = m_{X(3872)}$ as the input data), and the threshold of 
the $\phi\psi$ decay is higher, so that $X^s(+)$ would be stable against OZI-rule-allowed decays, 
when the mass value of $X^s(+)$ estimated above is taken literally, although it still has large  
uncertainties. 
On the other hand, $X^s(-)$ with $\mathcal{C}=-$ couples naturally to $D_s^{*+}D_s^{*-}$ which has 
$\mathcal{C}=-$ in the lowest ($S$-wave) state with a total angular momentum $J=1$. 
However, the decay $X^s(-)\rightarrow D_s^{*+}D_s^{*-}$ would not be allowed, because the estimated 
mass value, $m_{X^s(-)}\simeq 4.107$ GeV, which is obtained by using the same quark counting as 
the above (with $m_{X(-)} = m_{Z_c^0(3900)}$ as the input data), is lower than the  
threshold $2m_{D_s^{*}}\simeq 4.224$ GeV. 
Therefore, its main decays would be $X^s(-)\rightarrow \eta\psi$ and $\eta_c\phi$ with sufficient  
phase space volume.  

Next, in the same way as the above, we decompose  
$\tilde{X}^s(\pm) \sim 
\{[cs]_{\bar{3}_c}^{1_s}[\bar{c}\bar{s}]_{3_c}^{3_s}
\pm [cs]_{\bar{3}_c}^{3_s}[\bar{c}\bar{s}]_{3_c}^{1_s}\}_{1_c}^{3_s}$ 
and 
$\tilde{X}(\pm) \sim 
\{[cn]_{\bar{3}_c}^{1_s}[\bar{c}\bar{n}]_{3_c}^{3_s}
\pm [cn]_{\bar{3}_c}^{3_s}[\bar{c}\bar{n}]_{3_c}^{1_s}\}_{1_c}^{3_s}$ 
in the diquarkonium model. 
The results are  
\begin{eqnarray}                           
&&      \hspace{-6mm}
\tilde{X}^s(+) 
= \frac{1}{2}\sqrt{\frac{1}{3}}\Bigl\{
(\psi\phi +\phi\psi) - (D_s^{*+}D_s^{*-} + D_s^{*-}D_s^{*+)} 
\Bigr\} + \cdots,                           \label{eq:decomp-tilde-X^s_+}
\\
&&\hspace{-6mm}
\tilde{X}^s(-) 
= \frac{1}{2}\sqrt{\frac{1}{6}}\Bigl\{
(\eta_c\phi +\phi\eta_c) + (\psi\eta_s + \eta_s\psi)                                  \nonumber\\
&& \hspace{46mm}
- [D_s^+D_s^{*-} + D_s^{*+}D_s^- + D_s^-D_s^{*+} + D_s^{*-}D_s^+]
\Bigr\} + \cdots,                                                                    
\label{eq:decomp-tilde-X^s_-}
\\
&&\hspace{-5mm}
\tilde{X}(+) 
= \frac{1}{2}\sqrt{\frac{1}{6}}\Bigl\{
\sqrt{2}[\psi\omega + \omega\psi] 
- [D^{*0}\bar{D}^{*0} + \bar{D}^{*0}D^{*0} + D^{*+}{D}^{*-} + {D}^{*-}D^{*+}]
\Bigr\} + \cdots,                                                                       \label{eq:decomp-tilde-X(+)}
\\
&&\hspace{-5mm}
\tilde{X}(-) 
= \frac{1}{2}\sqrt{\frac{1}{6}}\Bigl\{
[\eta_c\omega + \omega\eta_c + \psi\eta_0 + \eta_0\psi] 
- [D^{0}\bar{D}^{*0} + \bar{D}^{0}D^{*0} + D^{*0}\bar{D}^{0} + \bar{D}^{*0}D^{0}]
\nonumber\\
&&\hspace{50mm}
- [D^{+}{D}^{*-} + {D}^{-}D^{*+} + D^{*+}{D}^{-} + {D}^{*-}D^{+}]
\Bigr\} + \cdots, 
\end{eqnarray}
where $\eta_0 = \{u\bar{u} + d\bar{d}\}_{1_c}^{1_s}/\sqrt{2}$. 
As seen in Eq.~(\ref{eq:decomp-tilde-X^s_+}),  $\tilde{X}^s(+)$ with $\mathcal{C}=+$ couples to not only 
$\phi\psi$ but also $D_s^{*+}D_s^{*-}$ with $\mathcal{C=-}$ in the lowest state of $J=1$.  
Such a coupling seems to be unnatural. 
On the other hand, as seen in Eq.~(\ref{eq:decomp-tilde-X^s_-}), $\tilde{X}^s(-)$ with $\mathcal{C}=-$  
couples to $D_s^+D_s^{*-} + c.c.$, in contrast with our model in which ${X}^s(+)$ couples naturally to 
$D_s^+D_s^{*-} + c.c.$ as seen above. 

In our previous work, rates for main decays of $X(-)$ and $X_I(-)=Z_c(3900)$ were very crudely  
calculated by taking a phenomenologically estimated rate 
$\Gamma(X(3872)\rightarrow D^0\bar{D}^{*0})\sim (0.3 - 1.5)$ MeV as the input data, 
and the resulting width of $X_I(-)$ was compatible with the measured ones of  
$Z^{\pm,0}_c(3900)$~\cite{Z_c-3900-KT}. 
However, we here revise the input data, and then the previous results on rates for decays of $X(-)$ 
and $X_I(-)$, and calculate rates for important decays of $X^s(\pm)$. 
Assuming that the full width of $X(3872)$ is approximately saturatated as 
$\Gamma_{X(3872)}\simeq \Gamma(X(3872)\rightarrow D^0\bar{D}^{*0} + c.c.) 
+ \Gamma(X(3872)\rightarrow\pi^+\pi^-\pi^0\psi) 
+ \Gamma(X(3872) \rightarrow \pi^+\pi^-\psi)$, 
and taking the measured ratios of decay rates,  
$\Gamma(X(3872)\rightarrow D^0\bar{D}^{*0} + c.c.)
/\Gamma(X(3872) \rightarrow \pi^+\pi^-\psi) 
= 9.5\pm 3.1$~\cite{Aushev},  
$\Gamma(X(3872)\rightarrow\pi^+\pi^-\pi^0\psi)
/\Gamma(X(3872) \rightarrow \pi^+\pi^-\psi) 
= 0.8\pm 0.3$~\cite{Choi-D0barDast0}, 
$\Gamma(X(3872) \rightarrow \gamma\psi)/\Gamma(X(3872) \rightarrow \pi^+\pi^-\psi) 
= 0.22\pm 0.09$~\cite{Z_c-3900-KT} 
(obtained by compiling the results from \cite{Belle-X-gamma-psi}) and the width  
$\Gamma_{X(3872)} = 3.9^{+2.8+0.2}_{-1.4-1.1}$ MeV~\cite{Aushev} (measured by using 
$X(3872)\rightarrow D^0\bar{D}^{*0} + c.c.$), as in \cite{Z_c-3900-KT}, we obtain phenomenologically 
$\Gamma(X(3872)\rightarrow D^0\bar{D}^{*0})_{\rm ph} = 0.81^{+0.72}_{-0.63}$ MeV   
and 
$\Gamma(X(3872)\rightarrow \gamma\psi)_{\rm ph} = 0.075^{+0.069}_{-0.061}$ MeV.  
The above $\Gamma(X(3872)\rightarrow D^0\bar{D}^{*0})_{\rm ph}$ is consistent with an independent  
estimate 
$\Gamma(X(3872)\rightarrow D^0\bar{D}^{*0})_{\rm Renga}\sim 1$ MeV~\cite{Renga}.    

Rate for the $X(+)\rightarrow D^0\bar{D}^{*0}$ decay was given in Eq.~(4) of \cite{Z_c-3900-KT}.  
Replacing  $X(+)$, $D^0$, $\bar{D}^{*0}$ by $X_I(-)$, $\pi^0$, $\psi$, respectively, we can obtain 
$\Gamma(X_I^0(-)\rightarrow\pi^0\psi)$ as 
\begin{equation}
\displaystyle{
\Gamma(X_I^0(-)\rightarrow \pi^0\psi) 
= \frac{|g_{X_I^0(-)\pi^0\psi}|^2}{24\pi m_{X_I(-)}^2}{p_\pi}
\biggl\{
2+ \frac{\bigl[m_{X_I(-)}^2 - m_{\pi^0}^2 + m_{\psi}^2\bigr]^2}
{4 m_{X_I(-)}^2m_{\psi}^2}
\biggr\}},                                                           \label{eq:rate-pi-psi}   
\end{equation}
where $p_\pi$ and $g_{X_I^0(-)\pi^0\psi}$ are the center-of-mass (c.m.) momentum of pion in the final  
state and the $X_I^0(-)\pi^0\psi$ coupling strength, respectively. 
In this way, the ratio 
$\Gamma(X_I^0(-)\rightarrow \pi^0\psi)/ \Gamma(X^0(+)\rightarrow D^0\bar{D}^{*0})$ 
is given by the ratio of coupling strengths ${|g_{X_I^0(-)\pi^0\psi}|}/{|g_{X^0(+)D^0\bar{D}^{*0}}|}$ 
which is given by the ratio of corresponding coefficients in the decompositions of $X_I(-)$ and $X(+)$ 
in Eqs.~(13)  and (10) of \cite{KT-partners-of-X}, when it is assumed that spatial wfs. of $X_I(-)$ and  
$X(+)$ are (nearly) equal to each other as in \cite{KT-partners-of-X}, and the $SU_f(4)$ symmetry  
breaking in spatial wfs. of ordinary mesons is ignored. 
Thus, $\Gamma(X_I^0(-)\rightarrow \pi^0\psi)$ can be revised by taking the above 
$\Gamma(X(3872)\rightarrow D^0\bar{D}^{*0})_{\rm ph}$ as the input data. 
In the same way, we revise rates for decays of $X(-)$ and $X_I(-)$, where the $\eta\eta'$ mixing with  
$\theta_P=-20^\circ$~\cite{PDG10} has been taken into account, and the  results are listed in Table~I. 
As seen in Eq.~(\ref{eq:decomp-X^s_-}), $X^s(-)$ couples to $\eta_c\phi$ and $\eta_s\psi$, so that 
it can decay into $\eta_c\phi$ and $\eta\psi$. 
Assuming again that spatial wfs. of $X^s(-)$ and $X(+)$ are (nearly) equal to  
each other, we can calculate very crudely rates for the $X^s(-)\rightarrow \eta_c\phi$ and $\eta\psi$  
decays, as listed in Table~I. 
It is seen that detection of $X^s(-)$ in $B$ decays would be difficult at the present accuracy, 
because it is expected to be considerably broad. 
This seems to be compatible with the result that no signal of narrow $\eta\psi$ peak has been 
observed in $B$ decays~\cite{Miyabayashi}. 
As for $X^s(+)$ in our model, it would have no OZI-rule-allowed decay, as discussed before. 
Because OZI-rule-suppression would be strong at the energy scale under consideration, its radiative 
decays would be important.    
Therefore, we calculate rates for the  
$X(+)\rightarrow\omega\psi\rightarrow\gamma\psi$, 
$X^s(+)\rightarrow\phi\psi\rightarrow\gamma\psi$ and  
$X^s(+)\rightarrow\phi\psi\rightarrow\phi\gamma$ decays  
under the vector meson dominance hypothesis~\cite{VMD} with photon-vector meson coupling  
strengths~\cite{VMD-Terasaki}  
$X_\rho(0) = 0.033\pm0.003$ GeV$^2$, $X_\omega(0) = 0.011\pm0.011$ GeV$^2$, 
$X_\phi(0) = -0.018\pm0.004$ GeV$^2$ and $X_\psi(0) = 0.050\pm0.013$ GeV$^2$, 
where their signs have been determined by using the quark model. 
(This approach reproduced well~\cite{omega-rho-KT} the measured ratio of decay rates 
$\Gamma(X(3872)\rightarrow\gamma\psi)/\Gamma(X(3872)\rightarrow\pi^+\pi^-\psi)$ mentioned  
above, where the formula for the decay rate $\Gamma(X(3872)\rightarrow \gamma\psi)$ as  
an example of rates for $1^+\rightarrow \gamma V$, ($V=\rho,\,\omega,\,\phi,\,\psi$) has been  
provided in \cite{omega-rho-KT}.)  
Under this condition, the ratios of decay rates 
${\Gamma(X^s(+)\rightarrow\gamma\psi)}/{\Gamma(X(+)\rightarrow\gamma\psi)} \simeq 1.7$ and 
${\Gamma(X^s(+)\rightarrow\gamma\phi)}/{\Gamma(X(+)\rightarrow\gamma\psi)} \simeq 6.2$ 
can be obtained, when it is assumed that spatial wfs. of $X^s(+)$ and $X(+)$ are (nearly) equal to  
each other. 
Thus we crudely estimate the rates $\Gamma(X^s(+)\rightarrow\gamma\psi)$ and 
$\Gamma(X^s(+)\rightarrow\gamma\phi)$ as listed in Table I, taking 
$\Gamma(X(+)\rightarrow\gamma\psi)_{\rm ph}$ 
as the input data. 
\begin{center}
\begin{table}[t]       %
\begin{quote}
Table~I. Rates for OZI-rule-allowed decays of hidden-charm axial-vector 
mesons $X(-)$,  $X_I^0(-)$ and $X^s(-)$ and radiative decays of $X^s(+)$ are listed. 
Phenomenological rates 
$\Gamma(X(+)\rightarrow D^0\bar{D}^{*0})_{\rm ph} = 0.81^{+0.72}_{-0.63}$ MeV and  
$\Gamma(X(+)\rightarrow\gamma\psi)_{\rm ph} = 0.075^{+0.069}_{-0.061}$ MeV 
which are given in the text are taken as the input data.
\end{quote} \vspace{2mm}
\hspace{25mm}
\begin{tabular}{|c|c|c|c|}
\hline
Decay & Rate (MeV)  & Decay & Rate (MeV) \\
\hline
$X(-)\rightarrow \eta_c\omega$  & \hspace{-7mm} $38^{+34}_{-30}$ 
&  
$X(-)\rightarrow\eta\psi$ 
& \hspace{-1mm}$24^{+22}_{-19}$ ($\ast$)   \\
\hline
$X_I^0(-)\rightarrow \eta_c\rho^0$  & \hspace{-7mm} $38^{+34}_{-30}$ 
&  
$X_I^0(-)\rightarrow\pi^0\psi$ 
& \hspace{-6mm}$58^{+51}_{-45}$    \\
\hline
\hspace{-3mm}
$X^s(-)\rightarrow \eta\psi$  &\hspace{-2mm} $20^{+18}_{-15}$  ($\ast$)
 & $X^s(-)\rightarrow \eta_c\phi$ &\hspace{-7mm} $33^{+29}_{-25}$ 
\\
\hline
\hspace{-3mm}
$X^s(+)\rightarrow \gamma\psi$ &\hspace{1mm} $0.12^{+0.11}_{-0.10}$   
& $X^s(+)\rightarrow \gamma \phi$
& \hspace{2mm}$0.74^{+0.68}_{-0.62}$ \\
\hline
\end{tabular}  \vspace{2mm}
\\
\hspace{40mm}($\ast$) $\eta\eta'$ mixing with $\theta_P= -20^\circ$~\cite{PDG10}
\end{table} \vspace{-5mm}
\end{center}

In summary we have compared the diquarkonium model with our simple tetra-quark model, and 
discussed that observation of double charm mesons will exclude the former. 
Then, decay properties of hidden-charm axial-vector mesons in these two different models have been 
compared with each other, by decomposing each of them into a sum of products of quark-antiquark  
pairs. 
As seen in the decompositions, our simple tetra-quark model can reproduce all the observed decay 
modes of $X(3872)$. 
It should be noted that the isospin non-conserving 
$X(+)=X(3872)\rightarrow (\rho^0\psi\rightarrow)\pi^+\pi^-\psi$ decay 
can proceed through the $\omega\rho^0$ mixing~\cite{omega-rho-KT}, although there is no direct  
$X(+)\rho^0\psi$ coupling. 
Its partners have necessary couplings to ordinary low lying mesons, but no unnatural one. 
On the other hand, axial-vector mesons in the diquarkonium model, in which the isospin 
symmetry has been assumed in this note (in contrast with the original one), have not only natural 
couplings but also unnatural ones, and lack a part of natural couplings, in particular, the iso-singlet  
$\tilde{X}(+)$ which could be assigned to $X(3872)$ does not couple to $D^0\bar{D}^{*0} + c.c.$ but 
its opposite $\mathcal{C}$-parity partner $\tilde{X}(-)$ does. 
It would imply that, in this model, $\tilde{X}(+) = X(3872)$ can decay into 
$(\omega\psi)\rightarrow 3\pi\psi$ and $(\omega\psi\rightarrow\rho^0\psi)\rightarrow 2\pi\psi$  
(through the $\omega\rho^0$ mixing) but not directly into $D^0\bar{D}^{*0} + c.c.$ 
In this case, the observed narrow $D^0\bar{D}^{*0} + c.c.$ peak at $3872$ MeV would come from  
$\tilde{X}(-)$ which is degenerate with $X(3872) = \tilde{X}(+)$. 
If so, a narrow $\eta\psi$ peak which is created in the $\tilde{X}(-)\rightarrow \eta\psi$ decay 
should be observed just at $3872$ MeV. 
However, it contradicts with the experiment~\cite{Miyabayashi} in which no signal of narrow $\eta\psi$  
peak has been observed. 
Contrary, in our tetra-quark model, it is expected that two broad $\eta\psi$ peaks which come from 
$X(-)\rightarrow\eta\psi$ and $X^s(-)\rightarrow\eta\psi$ will be observed 
in future.  
(Their detection in $B$ decays is probably difficult at the present experimental accuracy, 
because they are expected to be considerably broad.) 

Thus, experimental studies of decay properties of hidden-strangeness partners of $X(3872)$ 
will select a realistic model. 

\section*{Acknowledgments}    
The author would like to thank Professor T.~Hyodo for valuable discussions. He also would like to 
appreciate Professor H.~Kunitomo for careful reading of the manuscript. 



\begin{thebibliography}{99}
\setlength{\itemsep}{2pt} 

{

\bibitem{BESIII}
M.~Ablikim et al., BESIII Collaboration, arXiv:1303.5949 [hep-ex]. 

\bibitem{Belle-Z(3895)}
Z.~Q.~Liu et al., Belle Collaboration, arXiv:1304.0121 [hep-ex].

\bibitem{Xiao}
T.~Xiao, S.~Dobbs, A.~Tomaradze and K.~K.~Seth, arXiv:1304.3036 [hep-ex].  

\bibitem{Maiani-X}
L.~Maiani, F.~Piccinini, A.~D.~Polosa and V.~Riquer, Phys. Rev. D {\bf 71}, 014028 (2005). 

\bibitem{X-3872-KT}         
K.~Terasaki, Prog. Theor. Phys. {\bf 118}, 821 (2007); arXiv:0706.3944 [hep-ph].

\bibitem{Z_c-models}
Q.~Wang, C.~Hanhalt and Q.~Zhao, arXiv:1303.6355 [hep-ph]. 

\bibitem{Guo}
F.-K.~Guo, C.~Hidalgo-Duque, J.~Nieves, M.~Pavon and Valderrama, arXiv:1303.6608 [hep-ph].

\bibitem{Voloshin}
M.~B.~Voloshin, arXiv:1304.0380. 

\bibitem{Maiani-odd-C} 
R.~Faccini, L.~Maiani, F.~Piccinini, A.~Pilloni, A.D.~Polosa and V.~Riquer, 
arXiv:1303.6857 [hep-ph]. 

\bibitem{Z_c-3900-KT}              
K.~Terasaki, arXiv:1304.7080 [hep-ph].

\bibitem{Belle-X-3875}             
G. Gokhroo et al., Belle Collaboration, Phys. Rev. Lett. {\bf 97}, 162002 (2006).  

\bibitem{Babar-X-3875}
B. Aubert et al., Babar Collaboration, Phys. Rev. D {\bf 77}, 011102 (2008). 

\bibitem{prompt-X-theor} 
C.~Bignamini, B.~Grinstein, F.~Piccinini,A.~D.~Polosa, and C.~Saballi, Phys. Rev. Lett. 
{\bf 103}, 162001 (2009); arXiv:0906.0882 [hep-ph]; PoS EPS-HEP 2009:074, 2009. 
 
\bibitem{TWQCD} 
T.-W.~Chiu T.-H.~Hsieh, hep-ph/0603207. 

\bibitem{phi-psi-exp}                                                     
T.~A.~Aaltonen et al., CDF Collaboration, Phys. Rev. Lett. {\bf 102} , 242002 (2009);  
S.~Chatrchyan et al., CMS Collaboration, arXiv:1309.6920;                      
T.~V.~Uglov, Belle Collaboration, in {\it Proc. of 16th Int. Seminar on High Energy Physics 
(QUARKS 2010)};                                                                              
C.~P.~Shen et al., Belle Collaboration, Phys. Rev. Lett. {\bf 104}, 112004 (2010); 
R.~Aaij  et al., ${\rm LHC_b}$ Collaboration, Phys. Rev. D {\bf 85}, 091103  (2012). 

\bibitem{Miyabayashi}                                                                  
T.~Iwashita et al., Belle Collaboration, arXiv:1310.2704.

\bibitem{Jaffe} 
R.~L.~Jaffe, Phys. Rev. D {\bf 15} (1977), 267 and 281. 

\bibitem{D_{s0}-KT} 
K.~Terasaki, Phys. Rev. D {\bf 68} (2003), 011501(R);  
AIP Conf. Proc. {\bf 717}, 556 (2004);  hep-ph/0309279.    

\bibitem{HT-isospin}                
A.~Hayashigaki and K.~Terasaki, Prog. Theor. Phys. {\bf 114} (2005), 1191; hep-ph/0410393; 
K.~Terasaki, Invited talk at the workshop on {\it Resonances in QCD}, July 11 -- 15, 2005, 
ECT*, Trento, Italy; hep-ph/0512285.  

\bibitem{Belle-isospin}      
I.~Adachi et al., Belle Collaboration, arXiv:0809.1224 [hep-ex]; 
V.~Balagura, Belle Collaboration, Nucl. Phys. {\bf B} (Proc. Suppl.) {\bf 181}-{\bf 182}, 338 (2008). 

\bibitem{Babar-X-charged-partner}        
B.~Aubert et al., Babar Collaboration, Phys. Rev. D {\bf 71}, 031501 (2005).

\bibitem{Uehara}                                 
S.~Uehara et al., Phys. Rev. D {\bf 80}, 032001 (2009). 

\bibitem{hidden-charm-scalar-KT}         
K.~Terasaki, Prog. Theor. Phys. {\bf 121} (2009), 211; arXiv:0805.4460 [hep-ph].  

\bibitem{Exotic-QN} 
K.~Terasaki, Prog. Theor. Phys. {\bf 125}, 199 (2011); arXiv:1008.2992 [hep-ph];

\bibitem{Leutwyler}                             
H.~Leutwyler and M.~Roos, Z. Phys. C{\bf 25}, 91 (1984).

\bibitem{PDG96}                                 
R.~M.~Barnet et al., Particle Data Group, Phys. Rev.D {\bf 54}, 1 (1996)

\bibitem{FNAL-E687}                          
M.~S.~Nehring, Fermilab E687, Nucl. Phys. B (Proc.Suppl.) {\bf 55A}, 131 (1997). 

\bibitem{CLEO-FF}                             
J.~Barnett et al., CLEO, Phys. Lett. B{\bf 405}, 373 (1997).  

\bibitem{FF-SU-3-symm}
K.~K.~Seth et al., arXiv:1307.6587 [hep-ex]. 

\bibitem{OZI}
S.~Okubo, Phys. Lett. {\bf 5}, 165 (1963); G.~Zweig, CERN Report No. TH401 (1964); 
J.~Iizuka, K.~Okada and O.~Shito, Prog. Theor. Phys. {\bf 35}, 106 (1965). 

\bibitem{KT-partners-of-X}                  
K.~Terasaki, Prog. Theor. Phys. {\bf 127}, 577 (2012); arXiv:1107.5868 [hep-ph].

\bibitem{PDG10} 
K.~Nakamura et al., Particle Data Group, J. Phys. G {\bf 37}, 1 (2010). 

\bibitem{Aushev} %
T.~Aushev et al., Belle Collaboration, Phys. Rev. D {\bf 81}, 031103 (2010); arXiv:0810.0358. 

\bibitem{Choi-D0barDast0} 
S.-K.~Choi, hep-ex/1101.5691, and references quoted therein. 

\bibitem{Belle-X-gamma-psi}           
V.~Bhardwaj et al., Belle Collaboration, Phys. Rev. Lett. {\bf 107}, 091803 (2011); 
I.~Adachi et al., Babar Collaboration, arXiv:0809.1224 [hep-ex].

\bibitem{Renga}
F. Renga, Int. J. Mod. Phys. {\bf A26}, 4855 (2011); arXiv:1110.4151[hep-ph].

\bibitem{VMD} 
M.~Gell-Mann and F.~Zachariasen, Phys. Rev. {\bf 124}, 953 (1961). 

\bibitem{VMD-Terasaki} 
K.~Terasaki, Lett. Nuovo Cim. {\bf 31}, 457 (1981); Nuovo Cim. {\bf 66A}, 475 (1981). 

\bibitem{omega-rho-KT}    
K.~Terasaki, Prog. Theor. Phys. {\bf 122}, 1285 (2009); arXiv:0904.3368v2 [hep-ph]; 
Prog. Theor. Phys. Suppl. No.~{\bf 186}, 141 (2010); arXiv:1005.5573 [hep-ph]. 


}
\end{thebibliography}
\end{document}